# A fast sound power prediction tool for genset noise using machine learning.

Saurabh Pargal[a)], Abhijit A. Sane[b)]

**This paper investigates the application of machine learning regression algorithms—Kernel Ridge Regression (KRR), Huber Regressor (HR), and Gaussian Process Regression (GPR) for predicting sound power levels of gensets, offering significant value for marketing and sales teams during the early bidding process. When engine sizes and genset enclosure dimensions are tentative, and measured noise data is unavailable, these algorithms enable reliable noise level estimation for unbuilt gensets. The study utilizes high fidelity datasets from over 100 experiments conducted at Cummins Acoustics Technology Center (ATC) in a hemi-anechoic chamber, adhering to ISO 3744 standards. By using readily available information from the bidding and initial design stages, KRR predicts sound power with an average accuracy of ±5 dBA. While HR and GPR show slightly higher prediction errors, all models effectively capture the overall noise trends across various genset configurations. These findings present a promising method for early-stage noise estimation in genset design.**

Primary subject classification: enter primary; Secondary subject classification: enter secondary

## 1 INTRODUCTION

With the advent of the industrial revolution, fast technological advancement in various sectors, not only led humanity to a comfortable living but also an environment affected by its consequences; one of the major consequences is "Noise". For instance, gensets have become a necessity for industry, hospitals, malls, airports, and many other places as the main or standby source of power generation. But with it, it brings noise induced health hazards[1] such as health permanent hearing loss[2], annoyance, increased blood pressure[3], hypertension, disturbing sleep, and affecting cognitive functioning. To keep noise pollution in check, several organizations have set forward noise standards or certifications, to restrict noise levels (dB) to permissible levels. There are typically two categories of regulations that influence the level of noise exposure for individuals or the public: state or municipal noise ordinances and federal safety regulations established by the Occupational Safety and Health Administration (OSHA). The former pertains to noise that may extend beyond property boundaries and disrupt the public, although it is rarely loud enough to pose a safety risk. Conversely, the latter focuses on noise exposure standards within the workplace, aimed at safeguarding the health of employees.

Hence industries have implemented noise attenuation strategies as the part of the design and testing cycles in industries. But one of the main problems faced in implementing noise attenuation strategies into design, that it comes at the later stages of design/testing. And if the product exceeds the noise limitations, further design modifications to enclosures or other components are required, causing project delays.

a) Cummins Inc., Acoustics Technology Center (Fridley, MN, USA); email address: saurabh.pargal@cummins.com
b) Cummins Inc., Acoustics Technology Center (Fridley, MN, USA); email address: abhijit.a.sane@cummins.com

Therefore, this necessitates the need of a noise prediction tool at very early design stages, which can estimate sound power level, so that necessary design strategies can be implemented, to achieve noise levels as per regulations at final stages.

The issue at hand pertains to the prediction of noise levels at the genset-system level. The primary sources of noise include engine noise, which arises from mechanical and combustion processes, measuring between 100 to 121 dBA at a distance of one meter; cooling fan noise, characterized by dipole aeroacoustics, ranging from 100 to 105 dBA at the same distance; alternator noise, resulting from cooling air and brush friction, which falls between 80 to 90 dBA; induction noise, caused by fluctuations in the alternator windings, also within the 80 to 90 dBA range; engine exhaust noise, which can exceed 120 to 130 dBA without a silencer; and structural or mechanical noise, stemming from vibrations of various structural components. These noise sources arise from intricate physical phenomena influenced by numerous parameters. Furthermore, the design of the genset, including its enclosure, absorption materials, and component arrangement, varies significantly. Each genset product is distinct in terms of engine type, fan type, fuel type, and other characteristics.

Given the multitude of parameters and the complexity of the problem, a purely physics-based approach, such as finite element or boundary element simulations, is not feasible for predicting system-level noise. Researchers have explored simulation approaches, empirical methods based on ISO standards for predicting engine noise, fan noise, and overall genset noise for predicting engine noise, fan noise, and overall genset noise. For instance, when it comes to analytical approaches Bankar et al.[11], used statistical energy analysis for genset enclosure design, for noise prediction, More et al.[12] applied boundary element method for low frequency predictions for gensets, Arslan et al.[13] used finite element analysis for noise prediction for diesel gensets. These analytical models require design model, detail boundary conditions, noise source model or the noise source data itself for noise prediction. Most of these parameters are unavailable at preliminary stages and requires detail analysis. When it comes to diesel engine noise prediction, researchers in the past have used statistical approaches like independent component analysis[14], artificial neural networks[15] for prediction. But most of this research focused on component level noise prediction with a limited product range. Similarly, Askhedkar et al.[16] used artificial neural networks for genset noise prediction with several parameters, but study was limited to only one genset type. Researchers[17] have also used empirical approach to predict genset noise. Several approaches detailed above have been limited to specific components, such as a particular fan or engine, requires noise source data or to a narrow range of genset types with a limited parameter space. Considering the complexity of the problem and the numerous parameters involved, a data-centric approach is more practical. Over the past decade at Cummins, various genset products have been tested, yielding high-fidelity data, which supports the viability of a data-centric approach. Consequently, machine learning regression algorithms emerge as suitable candidates for predicting sound power levels. This methodology can also facilitate the modulation of fundamental parameters to optimize performance while minimizing noise levels, a process that can be implemented during the bidding stage.

At our hemi-anechoic chamber, several acoustics experiments have been conducted in last 10 years, for Cummin's gensets for various configuration, load conditions etc. Hence, a rich high-fidelity dataset on acoustics performance of several gensets is available. The objective of this study is to use this dataset and develop a sound power prediction tool based on machine learning algorithms: KRR[4], GPR[6], HR[5], to estimate sound power using very simple inputs available at initial design stages.

The organization of the paper is as follows. Section-2 describes the experimental approach used at our chamber to calculate sound power level of the genset, Section-3 discusses and presents the dataset used in this study, Section-4 details on the machine learning (ML) approach, Section -5 are focused on results and discussion on the performance of ML to predict sound power level and finally conclusion is presented in section-6.

## 2  Experiments: Sound power calculation of genset

Sound power level of a genset is calculated as per ISO 3744 standard, at Cummins hemi-anechoic chamber. In this approach, sound pressure levels are measured on an imaginary surface enveloping around the genset, approximating a free-field environment with only surface as reflecting plane, as sound is absorbed on all sides at Cummins' hemi-anechoic chamber.

The test procedure involves calibrating the microphones, acquiring the background noise for the chamber with unit off, running the genset for 15-20 minutes to reach stable condition, and then gathering data with unit on for several load conditions. For the noise data acquisition, a frequency bandwidth of 25600 Hz and a resolution of 1 Hz is used. The setup for a given genset is shown in Fig. 1 at chamber.

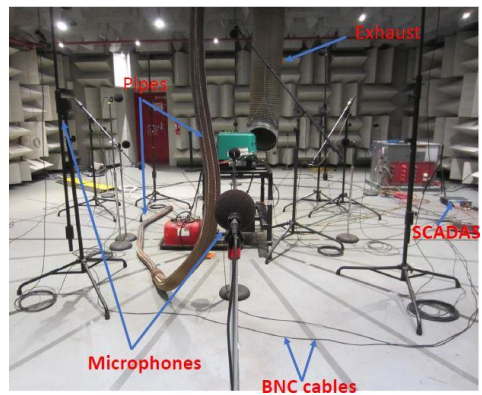
Fig. 1—Genset noise testing at Cummins hemi-anechoic chamber

Microphones are arranged in the near-field (at 1 meters) and far-field (at 3 or 7 meters) to measure sound pressure levels. For the near-field microphones, a parallel-piped arrangement is carried out to calculate sound power levels as per ISO 3744 standard, as represented in Fig. 2(a). Near-field and far-field arrangement of microphones around the genset, provides the noise directivity around the genset.

With this approach, different experiments were carried out with different configurations such as gensets with/without enclosures, various loads, with/without exhaust noise by connecting the exhaust to an outlet pipe etc. at Cummins noise testing facility. And sound power levels for several gensets is calculated for these different configurations.

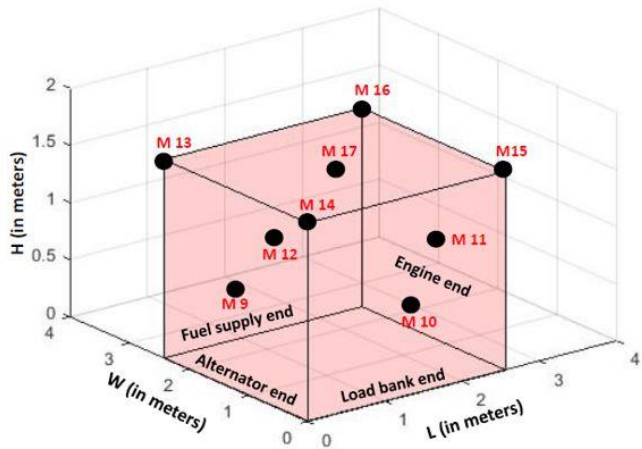
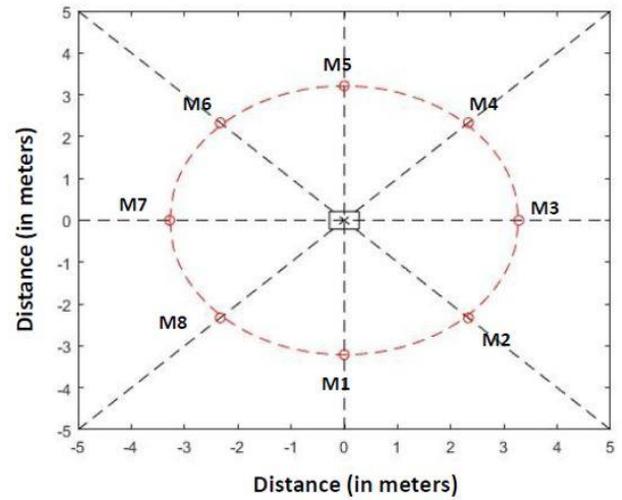

Fig. 2—Genset noise testing as per ISO 3744: (a) Microphone arrangement at 1 meter to calculate sound power level of genset (M9 to M17), (b) Microphone arrangement in the far-field (M1 to M8).

This standardized procedure to collect high-fidelity data following ISO 3744 standard, makes it an apt problem statement for regression algorithms in machine learning.

## 3    Dataset

As discussed previously, the problem statement is to predict sound power level at very initial design stages. Hence, only those dataset features are used that are available at initial bidding or design stages (see Fig. 3), such as: Engine, fuel type, overall estimated parallelopiped volume size of enclosure (with engine, generator, and other components), load (in k.W.), housing (whether it has an enclosure or not), and Engine RPM.

All these features are not only readily available at initial stages but also have a direct impact on sound power of the genset. For instance, larger engine load (in k.W.), leads to higher sound power. This is very clearly seen in the scatter matrix in Fig. 4, for Sound Power vs Load graph. Similar correlations can be seen for enclosure volume, and Engine RPM. These high correlations suggests that higher loads and rpm, these features will have dominant effect on sound power level prediction.

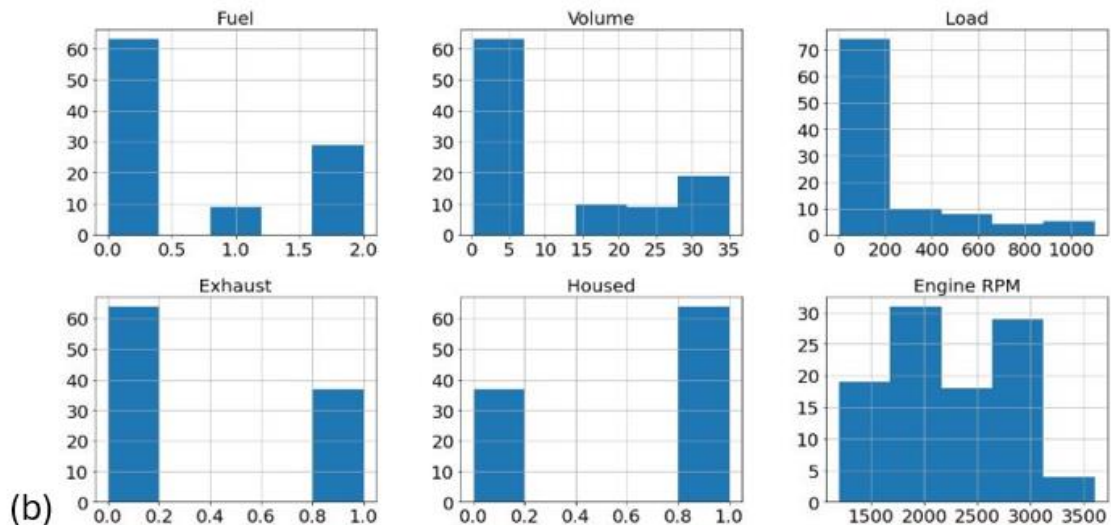

Fig. 3—Dataset: (a) Table (b) Histogram.

Next pair-plots are plotted in Fig. 5, using same features (Load, Volume, Engine RPM), showing dataset distribution for gensets with/without enclosure/housing (Fig. 5(a)), and with open/infinite exhaust (Fig. 5(b)). Similar correlations are seen between features, as observed previously in dataset scatter matrix. In our dataset, cases with open exhaust means exhaust noise is included in sound power calculation of the equipment, whereas in infinite exhaust, by the means of noise separation approach used within Cummins, exhaust noise is removed from the sound power calculation.

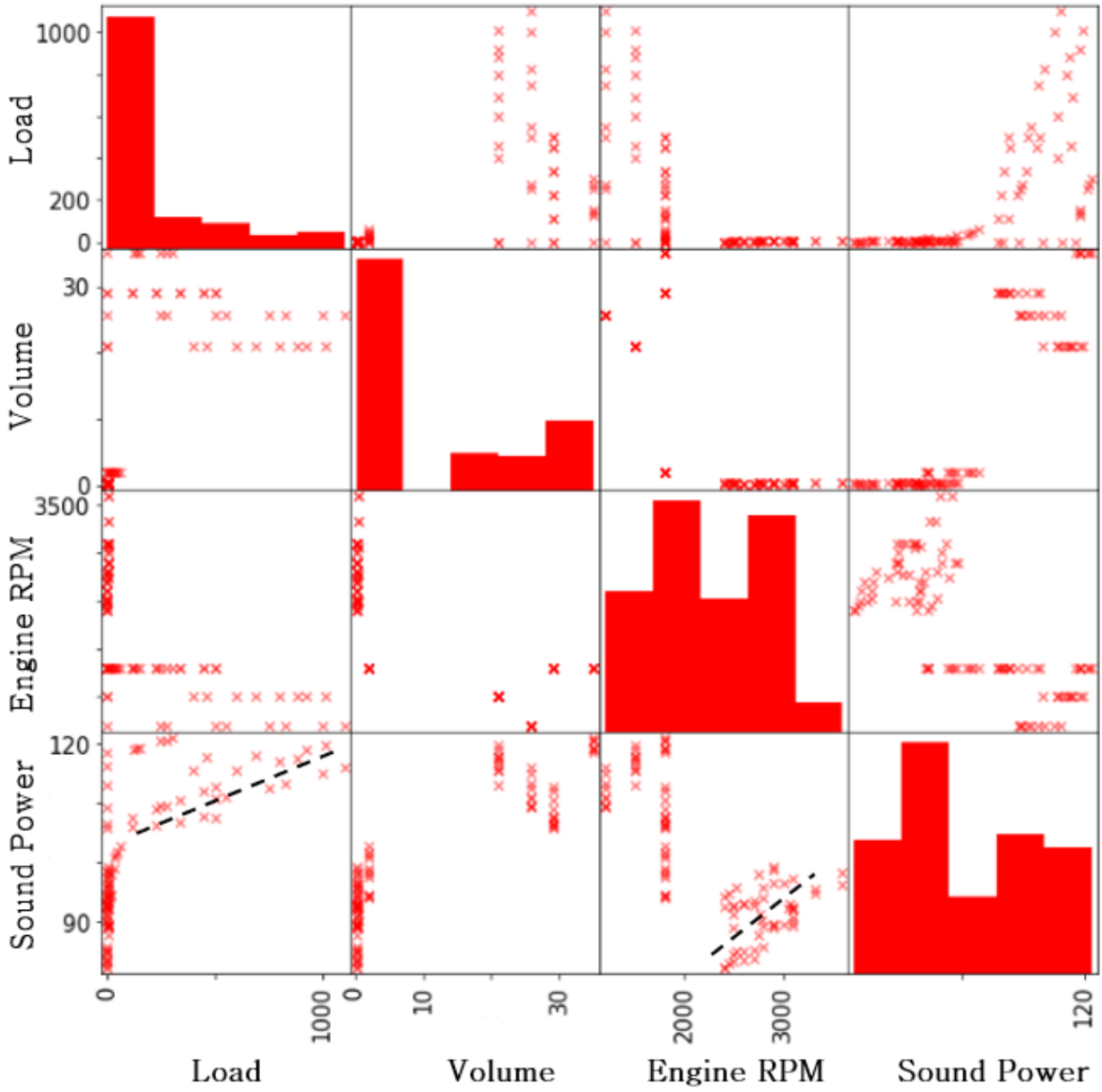

Fig. 4—Dataset scatter matrix.

In view of the strong correlations between sound power and load as seen in scatter matrix and pair plots, kernel density estimates are plotted in Fig. 6 for (a) open/infinite exhaust and (b) with/without enclosure. Interestingly, the plot reflects that the sound power scales with load much faster with open exhaust as compared to infinite exhaust. Hence that means, the sound power contribution from exhaust dominates with increasing load. Whereas in the case of gensets with enclosure, scaling with load is higher as compared to genset without enclosure, even though the sound power is higher in latter. This suggests that scaling of sound power with load, is directly affected by other parameters such as open/infinite exhaust or with/without enclosure.

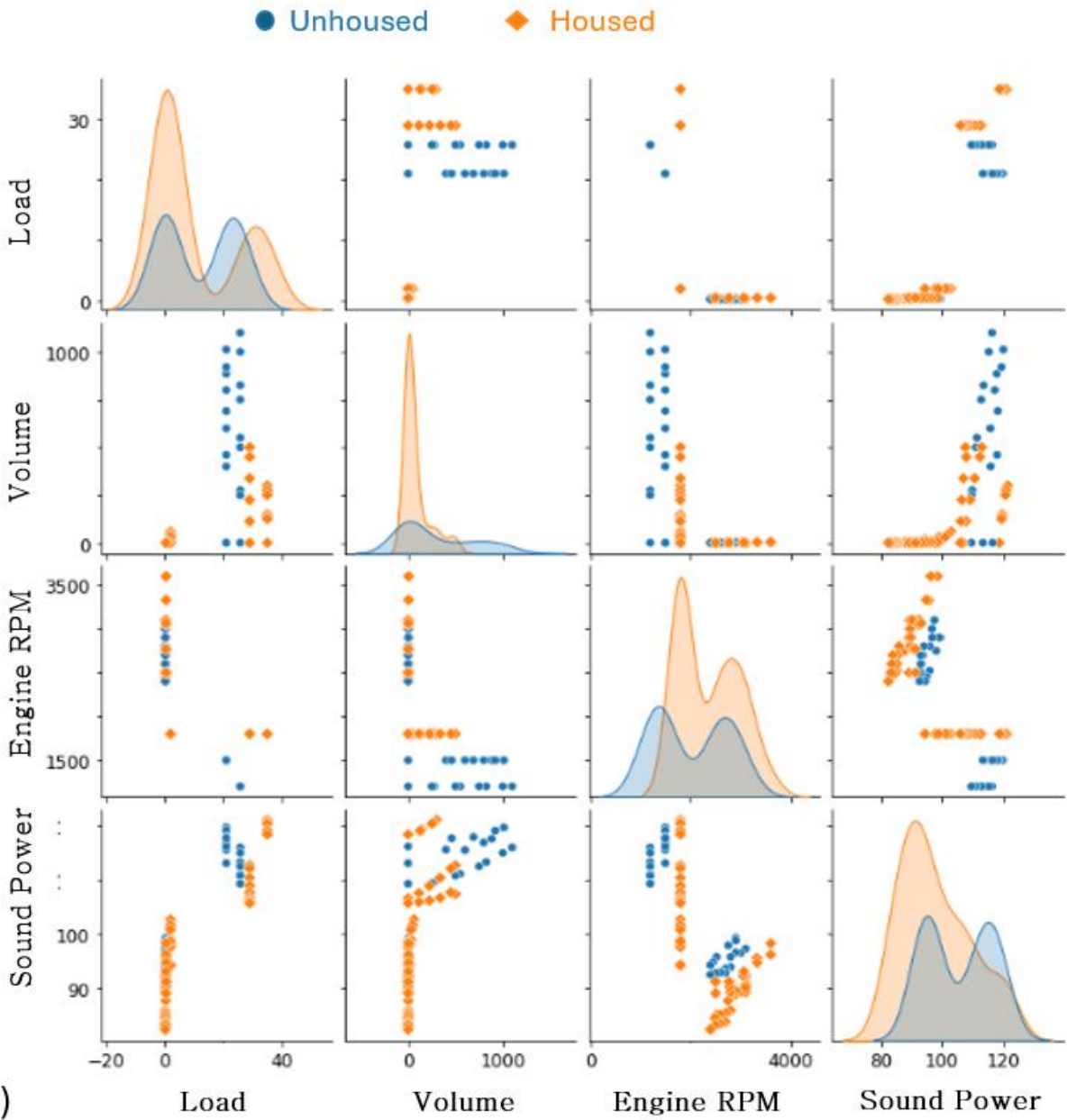

(a)

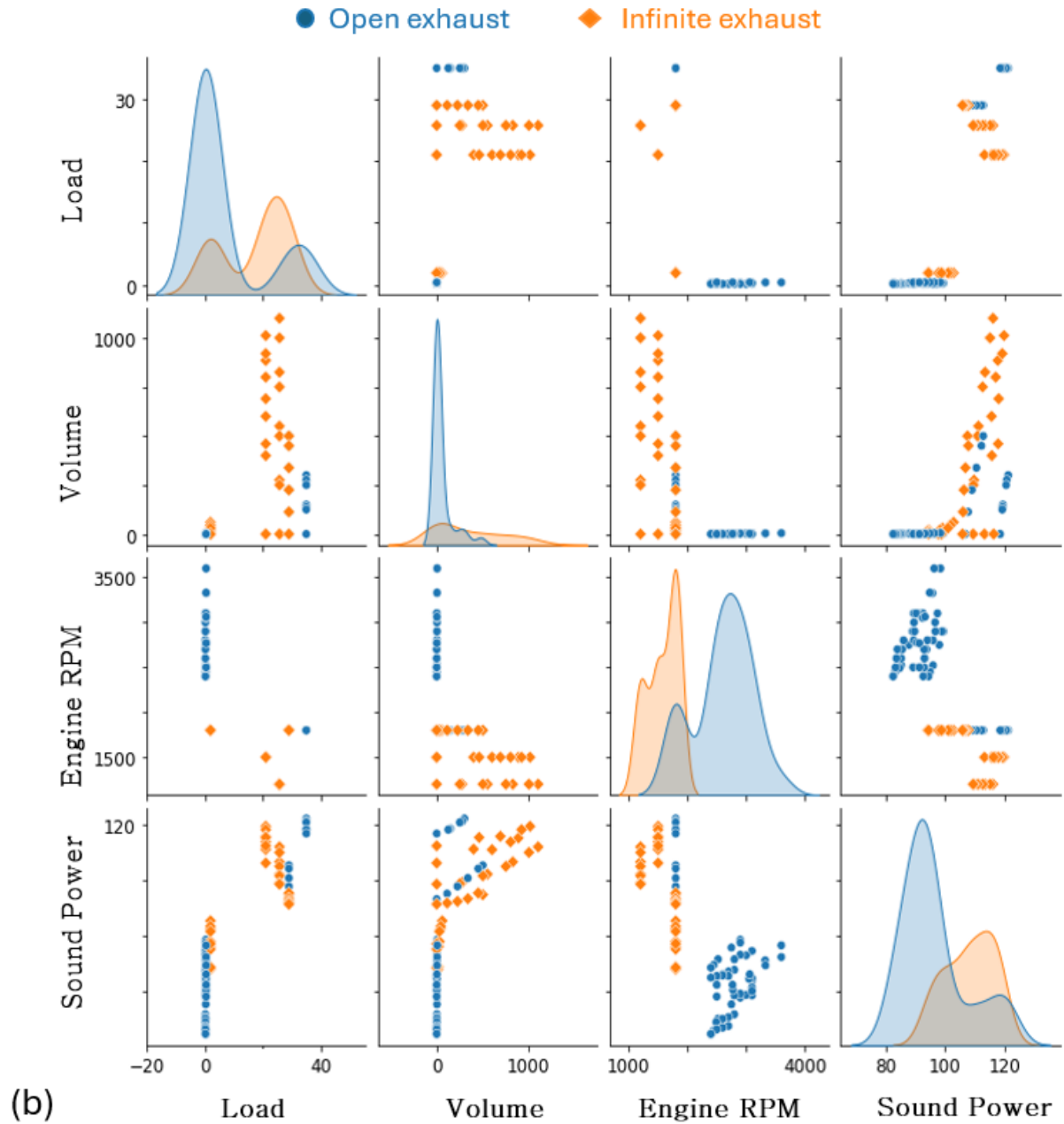

Fig. 5—Dataset pair-plot: (a) Enclosure with or without housing (b) Open (with exhaust noise) or infinite (removal of exhaust noise) exhaust.

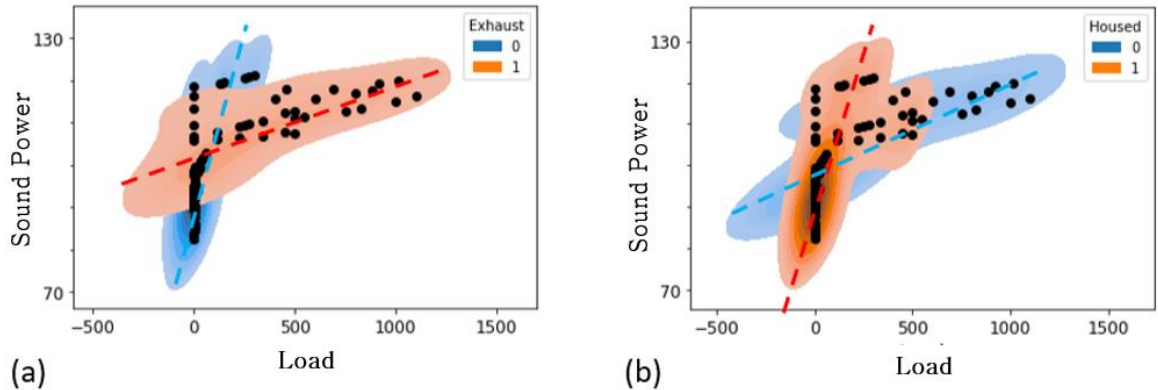

Fig. 6—Kernel-density estimate: (a) Sound Power vs Load, with (1) /without (0) exhaust noise (b) Sound Power vs Load with (1)/without (0) housing.

Next, further investigating the high correlation between sound power and load, a joint-distribution and kernel density estimate is plotted (Fig. 7), with also taking logarithmic of load. With logarithmic distribution of load, the correlation with Sound Power with load is even more clear. These findings suggests that machine learning regression algorithms are apt for sound power prediction with these features.

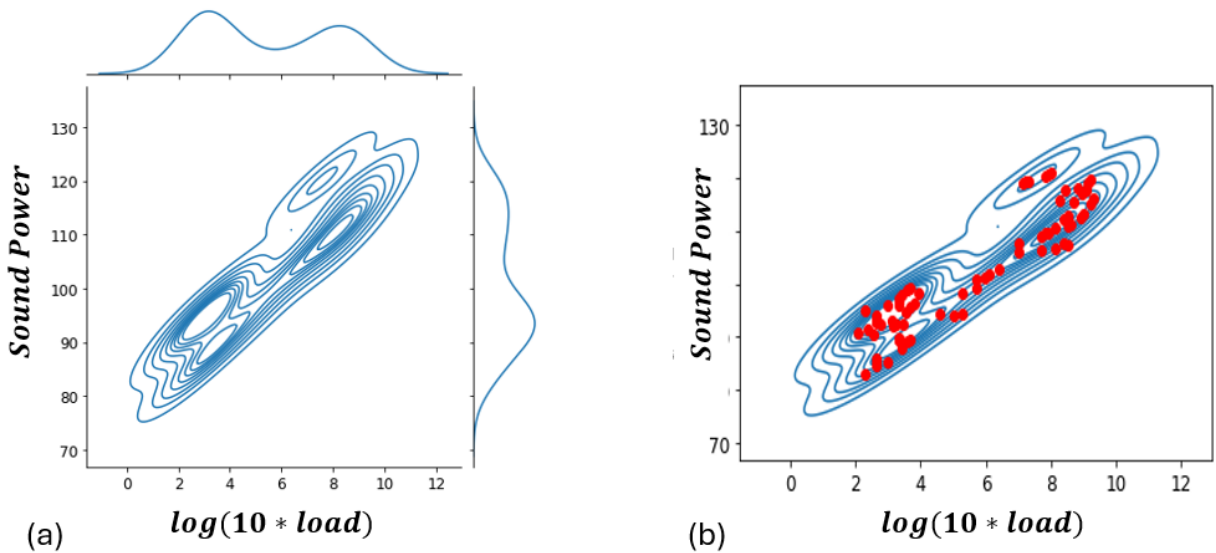

Fig. 7—(a) Joint distribution: Sound Power vs with logarithmic of Load. (b) Kernel density estimate with data points for Sound Power vs logarithmic of load.

## 4  Machine learning approach, dataset pre-processing and algorithms used.

Machine learning approach is majorly divided into following steps: 1. Data collection, 2. Data preparation (or pre-processing) for machine learning algorithms, 3. Splitting dataset into training and testing, 4. Training the ML algorithm on the training dataset and then finally testing the trained ML algorithm with testing dataset.

Dataset preprocessing is an important step as to correctly employ ML algorithms. This essentially involves cleaning dataset, dropping features or non-essential outliers. Such as certain outliers or data points were excluded from the dataset, including battery-powered or hybrid units that doesn't have a diesel engine, as well as units without a fan. This exclusion was necessary because diesel engines and fans are the primary sources of noise. And majority of dataset have diesel engine and fan, which directly effects overall noise. Datasets were also removed if they did not contain essential feature values, such as fan RPM and the presence or absence of exhaust, which significantly influence noise levels. Additionally, other characteristics, including enclosure width, type of absorption material, diesel engine type, number of cylinders, fan blade angle, and number of blades, were considered for exclusion for the following reasons: their impact on overall sound power was limited (<+/-5 dBA), feature values were not consistently available for all generators in the database, and such parameter information would not be accessible during the bidding process.

Next, dataset features such as enclosure housing (housed/unhoused) must be converted into a numerical equivalent i.e., for genset with enclosure- '1' and with no enclosure- '0'. Similarly other text labels such as with/without exhaust as well type of fuels (gasoline/natural gas/ diesel) are converted to numbers, so that they can be handled by algorithms. This is done using label encoders. In addition to numerical labeling, parameter features vary with different scaling. For instance, genset load varies from 0 to 1000, volume from 0 to 30, engine rpm up till 4000 etc. The variability of parameter space is scaled using standard-scaler, so that regression algorithms perform well.

Next, dataset is split into training and testing dataset (70-30 split between training and testing), by randomly selecting through the genset cases. The dataset includes generator set loads ranging from 2 kW to more than 1 MW, featuring various generator set sizes and configurations, such as those with or without exhaust systems and enclosures. Due to the diverse dataset, a stratified sampling method was employed for the test split. This approach ensures that for each generator set type, a portion of the data is allocated to both the training and testing phases. Consequently, machine learning models are exposed to some instances of each generator set during training prior to evaluation on the test data.

This study presents the performance of three distinct algorithms: Kernel Ridge Regression, Huber Regression, and Gaussian Process Regression (GPR). Kernel Ridge Regression is an extension of ridge regression that incorporates the kernel trick. Given that the parameter space is nonlinear, and the data is randomly distributed, the kernel trick facilitates the transformation of data into a higher dimensional space, where ridge regression is employed to mitigate the risk of overfitting. Subsequently, the Huber regressor utilizes a hybrid loss function that combines the quadratic loss function of linear regression with the absolute loss function characteristic of robust regression. This approach minimizes quadratic loss for small errors while applying linear loss for larger errors, thereby addressing the influence of outliers. This methodology is particularly relevant for gensets with limited data, such as those utilizing natural gas as a fuel source and operating under substantial loads. Lastly, Gaussian Process Regression represents a collection of random variables, each characterized by a joint Gaussian distribution, and it provides predictions in the form of probability distributions. Other algorithms were tested as well, but not shown as these algorithms showed most accurate results for the given data, notwithstanding their simplicity.

Finally, to assess the model's performance, the Root Mean Squared Error (RMSE) was calculated. The RMSE is defined as the square root of the mean squared error and functions as the standard deviation of the discrepancies between the actual data values and the predicted data values. The selection of RMSE is advantageous as it is expressed in the same unit as the target variable—sound power, facilitating easy interpretation and comparison.

## 5 Results and discussion

The Machine learning algorithms are trained on the training data (70 cases) and then are used for sound power prediction against testing data (30 cases). To assure ML performance, several stratified random train-test data splits are carried out. In this paper, ML prediction against two random data splits is shown.

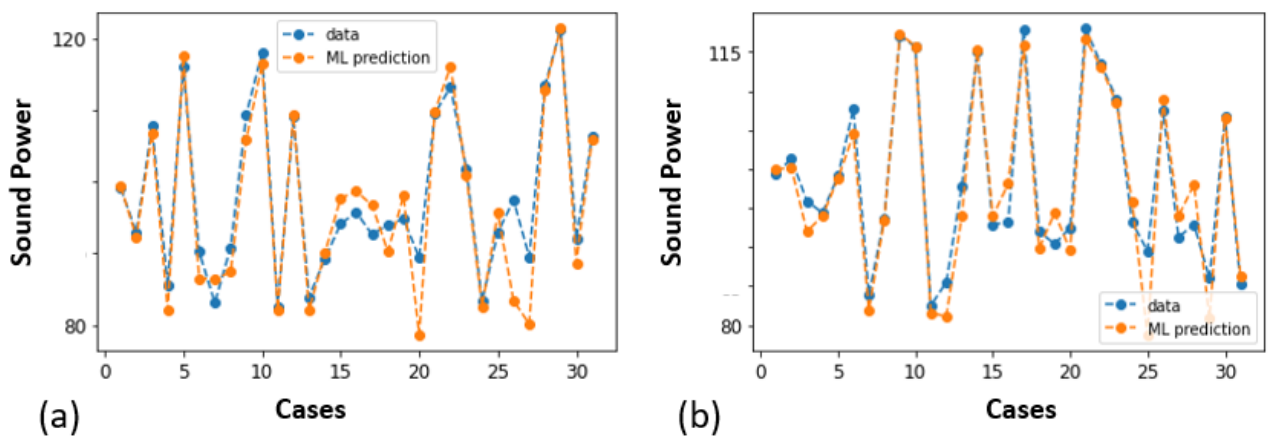

Fig. 8—Sound power prediction using Gaussian process regression (GPR) against test cases with random train-test splits (a) split 1 (b) split 2.

In Fig. 8, performance of GPR is evaluated against experimental data for different genset cases. Results indicate that the trend is captured well by GPR, but the average root mean square error (r.m.s.) is between 5 to 15 dB, across different train/test splits.

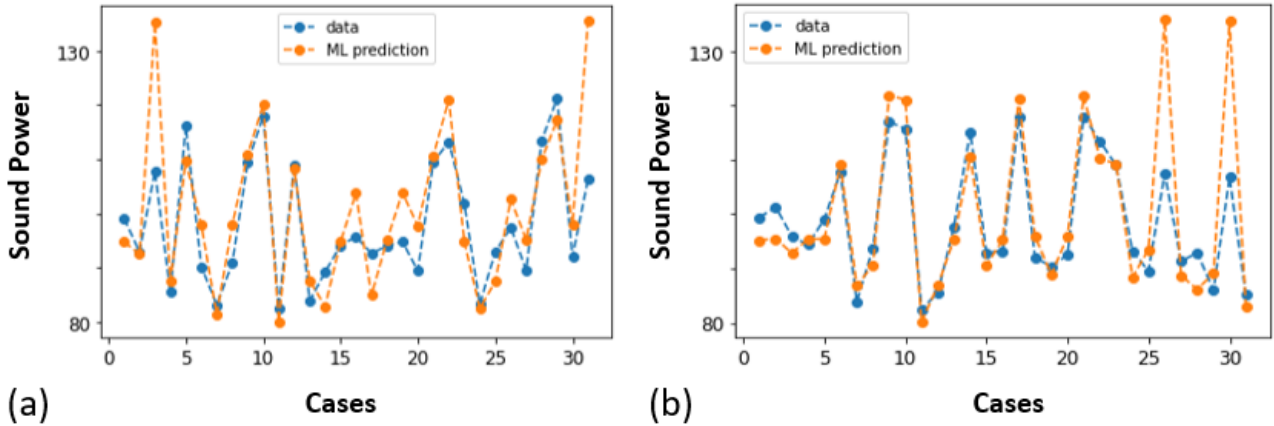

Fig. 9—Sound power prediction using Huber Regressor against test cases with random train-test splits (a) split 1 (b) split 2.

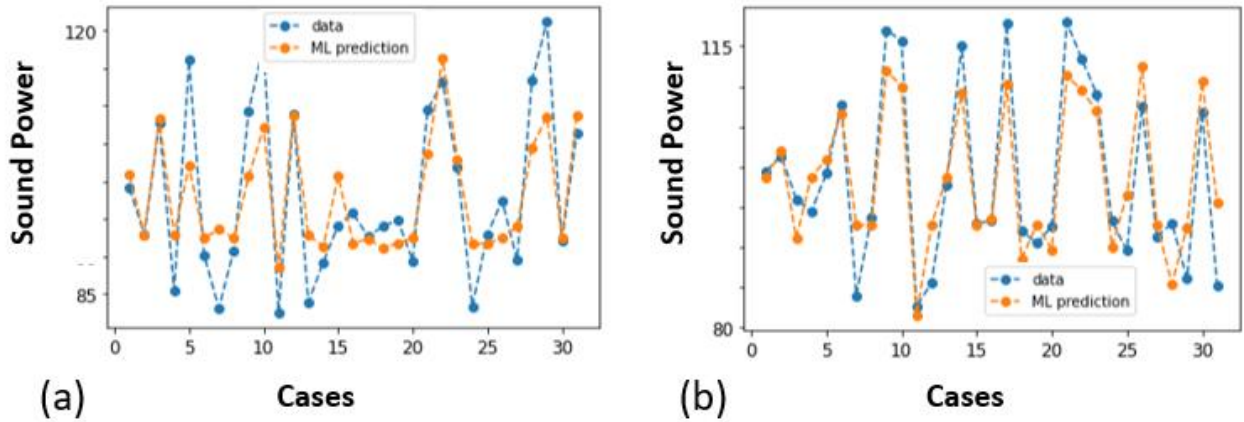

Fig. 10—Sound power prediction using Kernel Ridge Regressor against test cases with random train-test splits (a) split 1 (b) split 2.

Next, in Fig. 9, Huber Regressor performs better in comparison to GPR, across different train-test split. The average r.m.s. error is between 2 to 8 dB across different test/train splits. The higher error in some splits is attributed to extreme over or underprediction for some genset cases.

Finally in Fig. 10, Kernel Ridge regressor is tested which performs the best for the given dataset, with various splits. The average r.m.s. error is within 5 dB, and in some test/train split cases, it reaches less than 2 dB.

## 6 Conclusion

In this study, high fidelity dataset collected across several experiments conducted on Cummins gensets is used to predict sound power for gensets using machine learning algorithms, using simple features available at design/bidding stages.

Detail statistical analysis is carried out, to understand the variation of sound power with features considered such as load, rpm, enclosure size etc. Data preprocessing is carried out to employ ML algorithms accurately. Finally, ML algorithms such as KRR, GPR and HR are tested and showed overall good predictions against testing data, notwithstanding using only simple features. KRR performed the best across different test/training random splits, with accuracy reaching within +/- 5 dB.

In future work, the approach will be further employed to predict sound pressure levels in the near as well as far field of the genset. For more accurate noise predictions, more features will be added such as enclosure width or material (accounting transmission loss[7]), engine specifications (accounting combustion noise[8], vibroacoustics etc.), generator specifications, cooling fan blade number and rpm (fan noise[9,10]), to name a few.